\documentstyle[preprint,revtex]{aps}
\begin{document}
\draft
\preprint{UAHEP-9215}
\preprint{November 1992}
\begin{title}
Black Extended Objects, Naked Singularities and P-Branes
\end{title}
\author{B. Harms and Y. Leblanc}
\begin{instit}
Department of Physics and Astronomy, The University of Alabama \\
Box 870324, Tuscaloosa, AL 35487-0324
\end{instit}
\begin{abstract}
We treat the horizons of charged, dilaton black extended objects as quantum
mechanical objects.  We show that the
S matrix for such an object can be written
in terms of a p-brane-like action.  The requirements of unitarity of the S
matrix and positivity of the p-brane tension equivalent severely restrict the
number of space-time dimensions and the allowed values of the dilaton
parameter $a$.  Generally, black objects transform at the
extremal limit into p-branes.
\end{abstract}

\narrowtext

In Ref.\cite{cox} we analyzed the scattering of a massless, point-like
particle from a charged, dilatonic black hole \cite{garf},
assuming the horizon can be
regarded as a quantum mechanical object.
Following the approach of
 't Hooft \cite{thooft}, we showed that under rather general assumptions about
quantum gravitodynamics, the S matrix can be written in terms of a
``string-like'' action with an imaginary string tension
depending upon the dilaton parameter $a$.  For two values of $a\;(a =
\sqrt{3/5}, \sqrt{7})$ the string tension becomes positive
real beyond the extremal limit (coincidence of the
inner and outer horizons), and
black holes undergo a transition to strings in the region where,
classically, they would be naked singularities.  We proposed this as a new
way of implementing the ``cosmic censorship'' hypothesis and speculated that
such transitions could explain the observed extragalactic gamma ray
bursts, the gamma rays being the massless excitations of the strings
produced at the end point of the gravitational collapse.

In this paper we extend the work of Ref.\cite{cox} to higher dimensions by
considering black extended objects.  Horowitz and Strominger\cite{horo} have
obtained the metrics for charged black (10-D)-brane solutions of the field
equations which extremize the 10-dimensional action
\begin{eqnarray}
S = {1\over{16\pi}}\int d^{10}x \sqrt{-g}\bigl[ e^{-2\Phi}[R + 4(\partial
\Phi)^2] - {2\; e^{2\mu \Phi}\over{(D-2)!}} F^2\bigr]\; ,
\end{eqnarray}
where $\Phi$ is the dilaton field, R is the scalar curvature of the
10-dimensional space-time, and the field $F_{\mu_1...\mu_{D-2}}$ is a
(D-2)-form satisfying $dF = 0$.  The metrics for these black p-branes
were obtained
by requiring rotational and translational symmetry in (10-D) dimensions and
then substituting a general ten-dimensional metric with these symmetries into
Eq.(1) to obtain an effective D-dimensional action.  The goal of finding
black (10-D)-brane solutions of Eq.(1) is thus achieved by finding black hole
solutions to the D-dimensional action.  Gibbons and Maeda\cite{gibb}
have obtained the black hole solutions for such an action.  The most general
line element associated with static, spherically symmetric solutions to the
D-dimensional
equations of motion which are asymptotically flat and have a regular horizon
is of the form
\begin{eqnarray}
ds^2 = -\lambda^2 dt^2 + \lambda^{-2} d\hat{r}^2 + R^2 d\Omega_{D-2}^2,
\end{eqnarray}
where $\lambda$ and $R$ depend only on $\hat{r}$ and d$\Omega_{D-2}^2$ is the
line element on the unit (D-2)-sphere.  $\hat{r}$ is related to the usual
radius coordinate by $r^{D-4} dr = R^{D-4} d\hat{r}$, $\lambda^2$
is given by
\begin{eqnarray}
\lambda^2 = [1-({r_+\over{r}})^{D-3}]\;
[1-({r_{-}\over{r}})^{D-3}]^{1-\gamma (D-3)} \; ,
\end{eqnarray}
and $R^2$ is given by
\begin{eqnarray}
R^2 = r^2\; [1-(r_-/r)^b]^{\gamma}\; ,
\end{eqnarray}
with $b = D-3$.
The exponent $\gamma$ is given by
\begin{eqnarray}
\gamma = {2\; a^2 (D-2) \over{(D-3)[2(D-3)\; +\; a^2 (D-2)]}}\; ,
\end{eqnarray}
with
\begin{eqnarray}
a = \bigl[4\mu^2\; +\; 2\mu(7-D)\; +\;
2{D-1\over{(D-2)}}\bigr]^{1/2} \; .
\end{eqnarray}
The parameters $r_-$ and $r_+$ are respectively the inner and outer radii
of the horizons of a charged black hole.  They are related to the mass $M$
and charge $Q$ by
\begin{eqnarray}
M = {1\over{2}}\; [1-(D-3)\gamma]r_-^{D-3}\; +\; r_+^{D-3} \nonumber \\
Q = \bigl[{\gamma\; (D-3)^3\; (r_+r_-)^{D-3}\over{2a^2}}\bigr]^{1/2} \; .
\end{eqnarray}

To calculate the phase shift of a massless particle scattering from an
extended black object we follow the procedure of Ref.\cite{dray}.  The metric
of space-time surrounding such an object with a massless particle falling in
from spherical angle $\Omega' $ can be represented by gluing
together two solutions of Einstein's field equations.  The gluing is done
after a nonconstant shift of one of the Kruskal coordinates,
\begin{eqnarray}
\delta v(\Omega) = p_{in} f(\Omega, \Omega ')
\end{eqnarray}
along the null surface defined by setting the other Kruskal coordinate to
zero (u = 0).  The incident particle's momentum, $p_{in}$, is measured with
respect to the Kruskal coordinates and the Green function $f$ satisfies the
equation, at u = 0,
\begin{eqnarray}
{A\over{g}}\Delta f\; - {g_{,uv}\over{g}}f = 32\pi\; A^2\;
\delta^2(\Omega,\Omega')\big|_{u=0}\; ,
\end{eqnarray}
where $\Delta$ is the angular Laplacian in D-2 dimensions.  The Kruskal
coordinates are chosen such that the metric is of the form
\begin{eqnarray}
d\hat{s} = 2\; A(u,v) dudv\; +\; g(u,v)\; d\Omega^2_{D-2}\; .
\end{eqnarray}
The conditions for the matching of the two solutions to the field equations
after the shifting are
\begin{eqnarray}
A_{,v}|_{u=0} = g_{,v}|_{u=0} = 0\; .
\end{eqnarray}
The forms for $A(u,v)$ and $g(u,v)$ are determined by obtaining the
transformations from the $(\hat{r}, t)$ coordinates to the Kruskal
coordinates $(u,v)$.  This is accomplished by defining the ``tortoise''
coordinate $\xi$ as,
\begin{eqnarray}
\xi = \int {d\hat{r}\over{\lambda^2}}
\end{eqnarray}
with $\lambda^2$ being given by Eq.(3).  The Kruskal coordinates are obtained
as in, for example, Ref. \cite{adler}.  The form of the Kruskal coordinates
is the same as in the D = 4 case,
\begin{eqnarray}
u &=& e^{\alpha \xi}\;e^{\alpha t}\nonumber \\
& & \\
v &=& -e^{\alpha \xi}\; e^{-\alpha t}\;, \nonumber
\end{eqnarray}
where $\alpha$ is a constant whose value is determined by the matching
conditions (Eq.(11)).  These expressions for $u$ and $v$ give for $A(u,v)$ and
$g(u,v)$,
\begin{eqnarray}
A(u,v) &=& -{\lambda^2\;e^{-2\alpha \xi}\over{2\;\alpha^2}}\nonumber \\
& & \\
g(u,v) &=& R^2 \; . \nonumber
\end{eqnarray}
The vanishing of $A_{,v}$ and $g_{,v}$ at $u = 0\;(r = r_+)$ requires $\alpha$
to be,
\begin{eqnarray}
\alpha = {b\over{2\; r_+}}\; (1\;-\;(r_-/r_+)^b)^{1-\gamma (D-2)/2}\; .
\end{eqnarray}

The Green function $f$ which determines the phase shift of the particle being
scattered is determined by multiplying through Eq.(9) by $g$ and setting,
\begin{eqnarray}
\Gamma = {g_{,uv}\over{A}}\bigg|_{u=0} = {b\over{2}}\;(1-(r_-/r_+)^b)\bigl[
2\;+\;\gamma\;b\;(r_-/r_+)^b\;(1-(r_-/r_+)^b)^{-1}\bigr]\; .
\end{eqnarray}
To simplify the equation determining $f$ we take the ``north pole'' as the
direction of incidence, giving for $f$,
\begin{eqnarray}
\Delta f - \Gamma f = -2\; \pi\; \kappa\;\delta(\theta),
\end{eqnarray}
with
\begin{eqnarray}
\kappa = -16\;Ag\big|_{u=0} = {32\;e^{-1}\;
r_+^4\over{b^2}}\;[1-(r_-/r_+)^b]^{2\gamma -1}\; ,
\end{eqnarray}
where the constant of integration has been chosen such that this expression
for $\kappa$ agrees with that obtained for
the Schwarzschild black hole ($D = 4$ and $\gamma = 0$).

The solution to Eq. (17) has the same form as in \cite{cox}.  After expanding
the $\delta$-function in spherical harmonics and determining the expansion
coefficients, we find
\begin{eqnarray}
f = \kappa \sum_l\;{l+{1\over{2}}\over{l(l+1)+\Gamma}}\;P_l(\cos\theta)\; ,
\end{eqnarray}
where $\Gamma$ is as defined in Eq.(16) and $P_l(\cos\theta)$ is the $l$-th
order Legendre polynomial.

Having the function $f$ we can calculate the scattering matrix
\cite{thooft,dray},
\begin{eqnarray}
<u(\Omega)\mid v(\Omega)> = N\;\exp\biggl( i\int f^{-1}(\Omega,\Omega
')\;v(\Omega ')\;u(\Omega)\biggr)\; ,
\end{eqnarray}
where $N$ is a normalization constant and the propagator is given by
\begin{eqnarray}
f^{-1} = {(\Gamma\;-\;\Delta)\over{2\pi\;\kappa}}\; .
\end{eqnarray}
Fourier transforming the S matrix to the momentum representation we find
\FL
\begin{eqnarray}
<p_{out}(\Omega )\mid p_{in}(\Omega )\;>\; &=&\; N'\int\;Du(\Omega
)\;\int\;Dv(\Omega )\nonumber \\
&\times& \exp\biggl[
\int d^{D-2}\Omega\;\biggl({i\over{2\pi\;\kappa}}(\Gamma\;u\;v + \partial_{
\Omega}v\partial_{\Omega}u) + iu\;p_{out} - iv\;p_{in}\biggr)\biggr]\; .
\nonumber \\
\end{eqnarray}
The ``mass'' term ($\Gamma$ term), which is due to the curvature of the
horizon, will be neglected on the assumption that a patch can be
chosen small enough that it is locally flat.  To write Eq.(22) as a covariant
expression we introduce the coordinates
\begin{eqnarray}
x^0 = (v + u)/2\; ;\; x_{D-1} = (v - u)/2\; ,
\end{eqnarray}
where the metric $x^2 = {\bf x}^2 - x_0^2$ is used.  In terms of these
coordinates we can write Eq.(22) as,
\begin{eqnarray}
<p_{out}(\Omega) \mid p_{in}(\Omega)> &=&
C\int\;Dx^{\mu}(\sigma)\;Dg^{ab}\nonumber \\
&\times&
\exp\biggl[\int\;d^{D-2}
\sigma\;\biggl({-T\over{2}}[\sqrt{g}\;g^{ab}\;\partial_a
x^{\mu}\;\partial_b x^{\mu} + ix^{\mu}\;p^{\mu}(\sigma)]\biggr)\biggr]\; ,
\end{eqnarray}
where the integration $d^{D-2}\sigma$ is over variables
which are orthogonal to the membrane coordinates generated by
$x_0$ and $x_{D-1}$.  The string tension equivalent is given by
\begin{eqnarray}
T = {i\over{\pi\;\kappa}}\; .
\end{eqnarray}
The action term in Eq.(24) is of the Polyakov form except for the factor of
$i$, which prevents this S matrix element from satisfying unitarity.  To
overcome this difficulty we can analytically continue to the region in
parameter space where $r_- > r_+$ and require that the exponent $2\gamma -1$
in Eq.(18) satisfy the condition
\begin{eqnarray}
2\gamma - 1 = {n\over{2}}\; , \; n = odd \; ,
\end{eqnarray}
where $\gamma$ is given in Eq.(5).  The values of $a^2$ are thus
\begin{eqnarray}
a^2 = {2(D-3)^2\; (1 + n/2)\over{(D-2)(7-D) - (D-3)(D-2)\; n/2}}\; .
\end{eqnarray}
If we require $a$ to be real and the p-brane tension to be real and
positive, then only
$D = 4,5$ are allowed.  The values of $a$ for $D = 4$ were determined
in Ref.\cite{cox} to be $a = \sqrt{3/5}, \sqrt{7}$ for $n = 1, 5$
respectively.  For the $D = 5$ case only $n = 1$ is allowed, and the
corresponding value of the dilaton parameter is $a = 2$.
For these values of $a$ the constant, $T$, is real and
positive, and the action is that of a string theory for $D = 4$ and that of a
membrane theory in $D = 5$.

If we allow $T$ to be negative, then acceptable values of $a$ are obtained
for $D = 4,5,...,10$.  For $D>5$ however only $n = -1$ leads to acceptable
values of $a$.  For a string $T$ is the energy per unit length of the string.
So a negative value of $T$ does not seem to have any physical meaning.

If we
had carried through the curvature term in Eq.(22), we could have used the
condition $\Gamma = 0$, which gives the relation
\begin{eqnarray}
\biggl({r_-\over{r_+}}\biggr)^{D-3} = {a^2\;(D-2) + 2(D-3)\over{2(D-3)}}\; .
\end{eqnarray}
{}From Eq.(7) we also have
\begin{eqnarray}
\biggl({r_-\over{r_+}}\biggr)^b = {M^2\over{\delta^2}}{\biggl[1+\sqrt{1 -
{8a^2\;Q^2\delta^2\over{\gamma\;b^3\;M^2}}[1 -
\gamma\;b]}\biggr]^2\;\gamma\;b^3\over{8a^2\;Q^2[1- \gamma\;b]^2}}\; ,
\end{eqnarray}
where we have used the definitions
\begin{eqnarray}
b &=& D-3\; , \nonumber \\
\delta &=& {(D-2)\;\pi^{(D-3)/2}\over{8\Gamma({D-1\over{2}})}}\; .
\end{eqnarray}
Using Eqs.(28) and (29) we can obtain the ratio $Q/M$ in terms of $a$ and $D$
for which the curvature term vanishes.

As discussed in \cite{cox} the picture we obtain from the foregoing analysis
is that the charged, dilaton extended black objects avoid the
classical naked singularity for the case $r_- > r_+$ by undergoing a phase
transition at the extremal point $(r_- = r_+)$ and becoming p-branes for the
values of the dilaton parameter given above.

Our analysis shows that the results of Refs. \cite{cox,thooft} can be
extended to the case of charged, dilaton extended black objects.
Using the constraints of
unitarity and the positivity of the string tension, we find that the only
acceptable values of the dilaton parameter
$a$ are $a = \sqrt{3/5},\sqrt{7}$ for $D = 4$ and $a = 2$ for $D = 5$.
Since $D = 4,5$ are the only
allowed values of $D$, the only acceptable solutions to Eq.(1) are the black
6-brane and the black 5-brane, which undergo transitions into strings and
membranes respectively at the extremal limit.
The conditions of unitarity of
the S matrix and positivity of the p-brane tension put surprisingly strong
restrictions on the values of $D$ and $a$.  Furthermore duality in the
dilaton parameter
as defined in
Ref.\cite{cox} exists only in 4 dimensions.

\acknowledgments

This research was supported in part by the U.S. Department of Energy under
Grant No. DE-FG05-84ER40141.

\end{document}